\documentclass[
hf,
]{ceurart}

\usepackage{hyperref}
\usepackage[english]{babel}
\usepackage{graphicx}
\usepackage{subcaption}

\usepackage{hyperxmp}
\usepackage{listings}
\addto\extrasenglish{%
}
\begin{document}

\copyrightyear{2025}
\copyrightclause{Copyright for this paper by its authors.
  Use permitted under Creative Commons License Attribution 4.0
  International (CC BY 4.0).}

\conference{Workshop on
Recommender Systems for
Sustainable Development (RS4SD), co-located with CIKM'2025, November 10--14,2025, Seoul, Republic of Korea}

\title{SmartSustain Recommender System: Navigating Sustainability Trade-offs in Personalized City Trip Planning}

\tnotemark[1]
\author[]{Ashmi Banerjee}[%
email=ashmi.banerjee@tum.de,
url=https://ashmibanerjee.com,
]
\cormark[1]
\address[]{Technical University of Munich, Germany},

\author[]{Melih Mert Aksoy}[%
email=aksoy@in.tum.de,
]

\author[]{Wolfgang W\"orndl}[%
email=woerndl@in.tum.de,
]

\cortext[1]{Corresponding author.}

\begin{abstract}
Tourism is a major contributor to global carbon emissions and over-tourism, creating an urgent need for recommender systems that not only inform but also gently steer users toward more sustainable travel decisions. Such choices, however, often require balancing complex trade-offs between environmental impact, cost, convenience, and personal interests. 
To address this, we present the SmartSustain Recommender, a web application designed to nudge users toward eco-friendlier options through an interactive, user-centric interface. The system visualizes the broader consequences of travel decisions by combining CO\textsubscript{2}e emissions, destination popularity, and seasonality with personalized interest matching. It employs mechanisms such as interactive city cards for quick comparisons, dynamic banners that surface sustainable alternatives in specific trade-off scenarios, and real-time impact feedback using animated environmental indicators. A preliminary user study with 21 participants indicated strong usability and perceived effectiveness.
The system is accessible at \url{https://smartsustainrecommender.web.app/}.
\end{abstract}


\maketitle
\section{Introduction}
Online travel platforms have democratized access to global destinations; yet, their recommendation algorithms have traditionally optimized for user preferences and cost, largely ignoring the substantial environmental and societal costs associated with tourism~\cite{banerjee2023fairness}. As awareness grows around issues like climate change and over-tourism, there is an urgent need for tools that not only present sustainable options but also help users comprehend the complex trade-offs involved. 
A more sustainable travel choice may involve a higher price, a longer journey, or a destination that is a slightly weaker match for a user's stated interests. 

This paper introduces the \textbf{SmartSustain Recommender}, a demonstration system designed to address this challenge. It is a user-centric, modular, web application that operationalizes principles from decision science and visualization to make sustainability trade-offs transparent, manageable, and engaging. 
More importantly, the system embeds nudging mechanisms into the user interface: dynamic banners surface sustainable alternatives when users engage with high-impact options, positive reinforcements celebrate low-emission decisions, and interactive visualizations make the environmental consequences of choices tangible and intuitive.

Our work builds on existing research in sustainable TRS, which has sought to integrate environmental considerations into travel planning~\cite{banerjee2024modeling,banerjee2024sustainable}. However, many of these systems focus primarily on providing information about sustainable options without adequately addressing the cognitive and emotional barriers that users face when making travel decisions~\cite{banerjee2024green}. In contrast, the SmartSustain Recommender emphasizes user engagement through interactive visualizations and real-time feedback, making the trade-offs between sustainability and personal preferences more transparent and manageable. Additionally, our system extends the Green Destination Recommender~\cite{banerjee2024green} by providing a more personalized and context-aware approach to guiding users toward eco-friendly travel choices.
A preliminary user study with 21 participants further confirmed the system's usability and perceived effectiveness in promoting sustainability, while also highlighting opportunities to refine certain interface elements.

\textit{Contributions.}
The primary contributions of this work are centered on the system's design and implementation: (1) a novel, interactive interface that consolidates multi-dimensional data into a cohesive, at-a-glance dashboard for easy comparison; (2) the implementation of \textit{Dynamic Trade-Off Banners}, a context-aware nudging mechanism that proactively suggests alternative destinations to help users explore specific compromises (e.g., better sustainability for a slightly higher cost). (3) The integration of \textit{Real-Time Impact}, using visual metaphors like animated environmental indicators to make the abstract consequences of travel choices more salient and emotionally resonant.
The system, accessible at \url{https://smartsustainrecommender.web.app}, serves as a research platform to explore how different interface designs can influence eco-conscious decision-making.

The remainder of this paper is structured as follows. \autoref{section: related} reviews related work in sustainable tourism recommenders and user experience design for decision support. \autoref{section: user_journey} details the user journey, while~\autoref{section: implementation} covers system architecture, and preliminary validation through a user study. Finally, \autoref{section: future} discusses future research directions and concludes the paper.

\section{Related Work} \label{section: related}
Our work is situated at the intersection of sustainable tourism recommenders and user experience design for decision support, hence we review related literature in both domains.

\paragraph{Sustainable Tourism Recommenders}
Recent research has begun to integrate sustainability into tourism recommenders. The work of~\citet{banerjee2024modeling, banerjee2024green} forms the foundation for our system by proposing a framework to model sustainable city trips using CO\textsubscript{2}e emissions, popularity, and seasonality. Other research has focused on multi-stakeholder fairness, balancing the needs of travelers, local businesses, and residents to mitigate issues like over-tourism~\cite{abdollahpouri2021multistakeholder, merinov2024positive}.~\citet{mauro2024point} explored nudging users towards sustainable points-of-interest. Our work extends these efforts by focusing specifically on the \textit{communication of trade-offs} to the end-user, moving beyond algorithmic optimization to user-centric interaction design.

\paragraph{User Experience and Decision Support}
Making complex choices can lead to cognitive overload and decision fatigue~\cite{echterhoff2024avoiding}.~\citet{schnabel2016shortlists} demonstrated that tools like shortlists can reduce this burden by helping users manage options. We apply this principle by designing components that offload the mental effort of comparing multiple factors. Furthermore, research by~\citet{okoso2024tone} has shown that the tone of explanations can significantly impact user engagement. We incorporate this by using motivational and positive framing in our nudges, aiming to encourage sustainable choices rather than merely presenting data. The SmartSustain Recommender combines these user experience (UX) principles to create an environment where users do not feel burdened by too many choices.

\begin{figure}[htbp]
    \centering
    \includegraphics[width=0.6\linewidth]{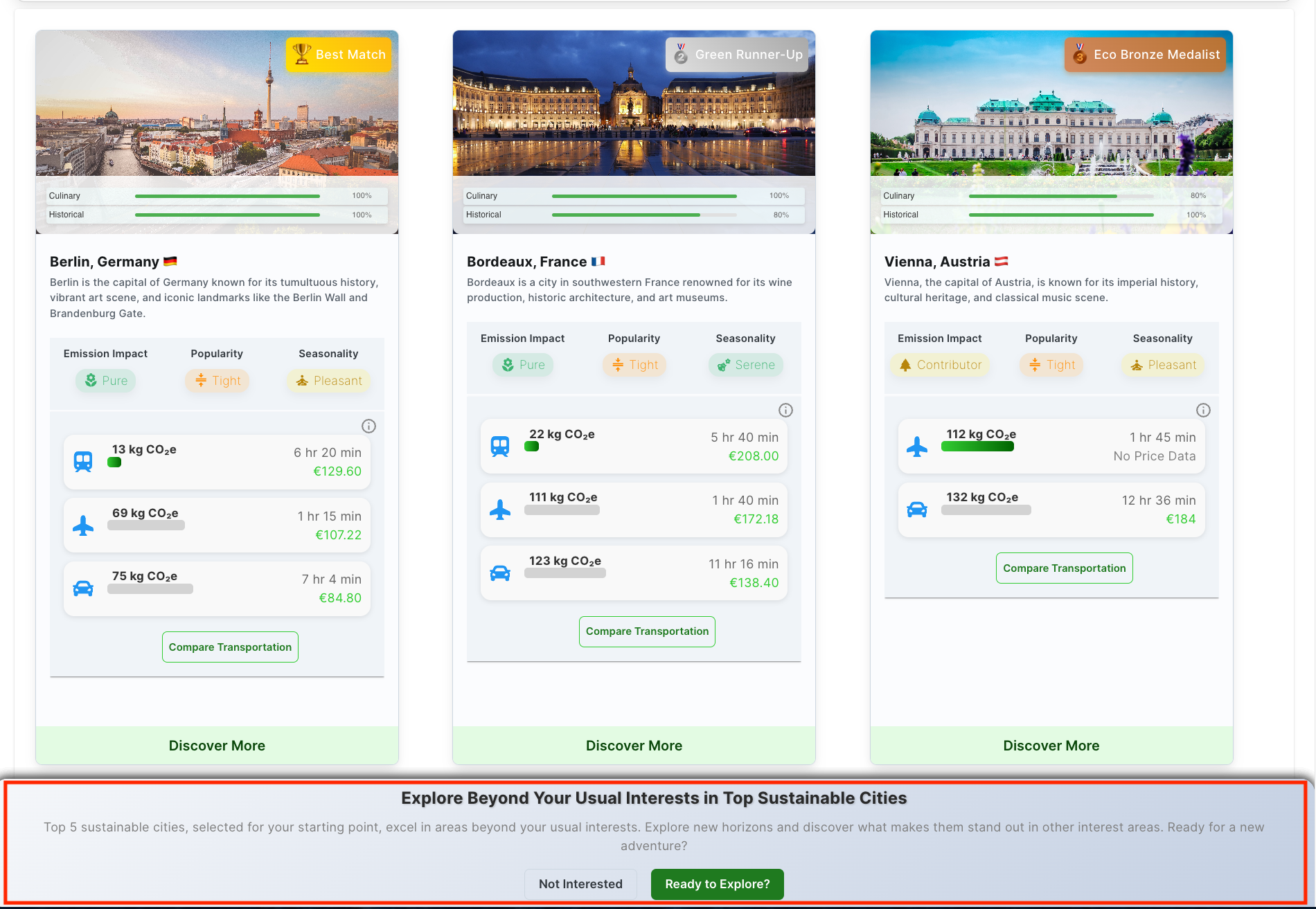}
    \caption{Card view with sustainability tags, transport summary, and interest alignment bars. The dynamic banner below, highlighted in \textcolor{red}{red}, aims to encourage users to make more sustainable choices.}
    \label{fig:card_view_single}
\end{figure}

\begin{figure}[htbp]
    \centering
    \includegraphics[width=0.6\linewidth]{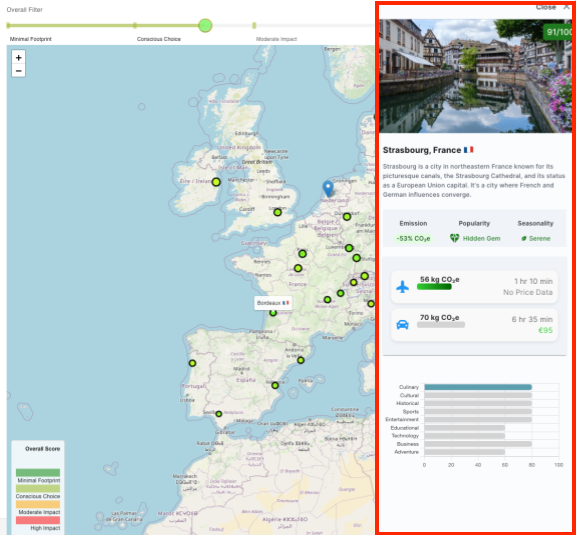}
    \caption{Map view with interactive markers and polylines representing destinations and transport options. Upon clicking on one of the markers, users can view detailed information about the destination and available transportation (highlighted in \textcolor{red}{red}).}
    \label{fig:map_view_single}
\end{figure}

\section{User Journey and Interface} \label{section: user_journey}
The SmartSustain Recommender is designed as an intuitive and interpretable system that guides users through the entire decision-making process, helping them balance sustainability, cost, popularity, and interest alignment.
The current use case focuses on European city travel, helping a traveler identify suitable destinations for leisure in a given month.

\paragraph{Landing Page}  
The journey begins on the \textit{Landing Page}, where users specify departure city, travel month, and interests (e.g., cultural, culinary, historical). An optional advanced survey captures finer-grained preferences such as walkability, nightlife, or climate vulnerability. A ranked list of destinations is then generated using a linear scoring function blending sustainability, interest alignment, popularity, and seasonality (\autoref{subsection: scoring}).

\paragraph{Explore Workspace}  
Recommendations are presented through two complementary views:  
\begin{itemize}
    \item \textbf{Card View} (\autoref{fig:card_view_single}): Compact dashboards display sustainability tags (emissions, popularity, seasonality), transport summaries, and interest alignment bars, supported by color-coded icons and badges.  
    \item \textbf{Map View} (\autoref{fig:map_view_single}): An interactive Leaflet map visualizes destinations with colored markers and polylines representing impact, popularity, and transport emissions. Threshold-based animations and consistent legends support rapid exploration. Clicking a destination reveals detailed trade-offs (CO\textsubscript{2}e, cost, duration, and interest fit) as highlighted in \textcolor{red}{red} in~\autoref{fig:map_view_single}.  
\end{itemize}

\begin{figure}[htbp]
    \centering
    \includegraphics[width=0.6\linewidth]{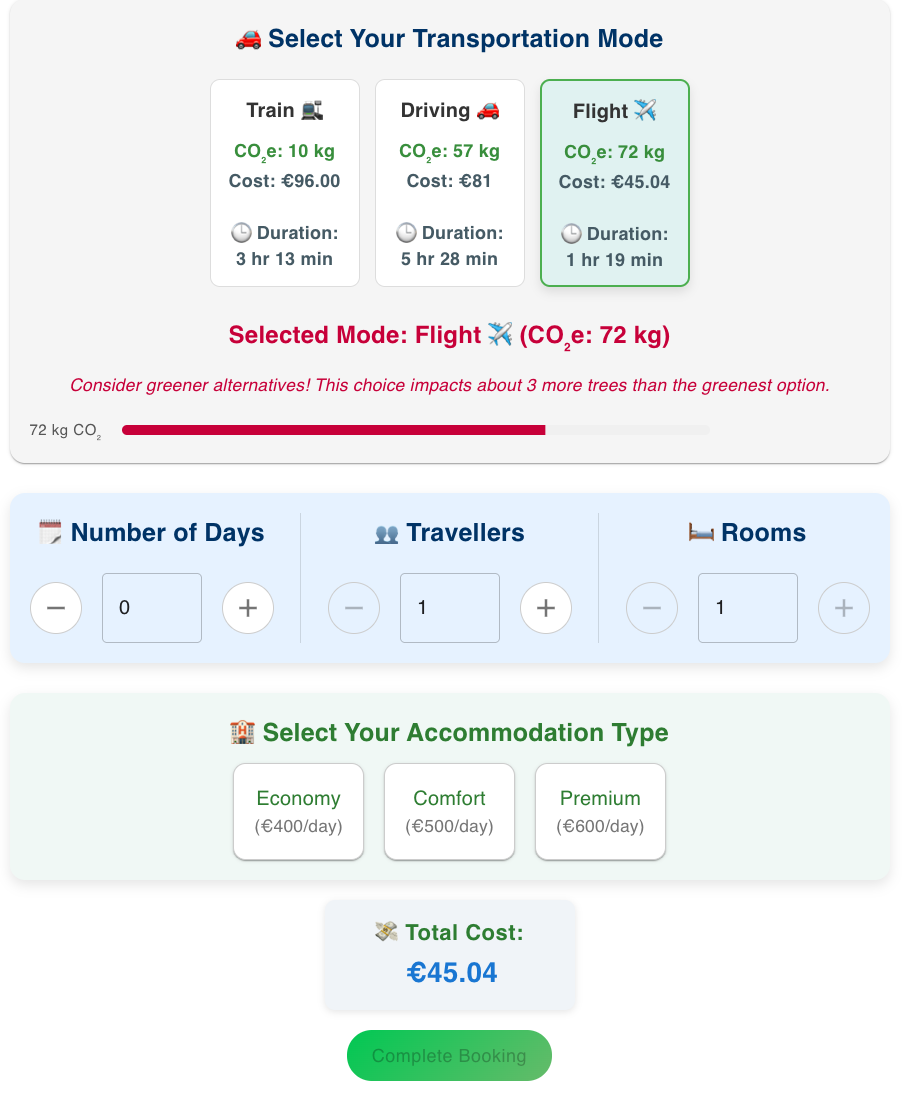}
    \caption{Booking component with transport and accommodation options showing financial and environmental impacts.}
    \label{fig:booking_page}
\end{figure}

\paragraph{City Details and Booking Flow}  
Selecting a city opens a \textit{details panel} consolidating climate vulnerability, transport options, and interest breakdowns. 
To compare the different modes of transport (e.g., train, bus, flight), users can access a \textit{comparison view} that highlights key metrics such as CO\textsubscript{2}e emissions, cost, and travel time through a radar chart.
Once the user finalizes their city, they can proceed to the booking page (\autoref{fig:booking_page}), where they can specify transport, accommodation, and group size. The interface displays both financial and environmental impact, simulating a realistic booking flow while reinforcing sustainability awareness.

\begin{figure}[htbp]
    \centering
    \includegraphics[width=0.8\linewidth]{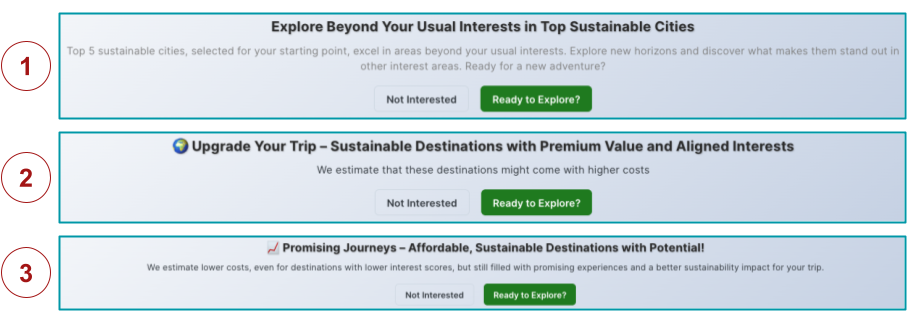}
    \caption{Different banners promoting sustainable travel options. They are contextually displayed based on user interactions and preferences to nudge them towards greener choices.}
    \label{fig:banners}
\end{figure}

\begin{figure}[htbp]
    \centering
    \includegraphics[width=0.8\linewidth]{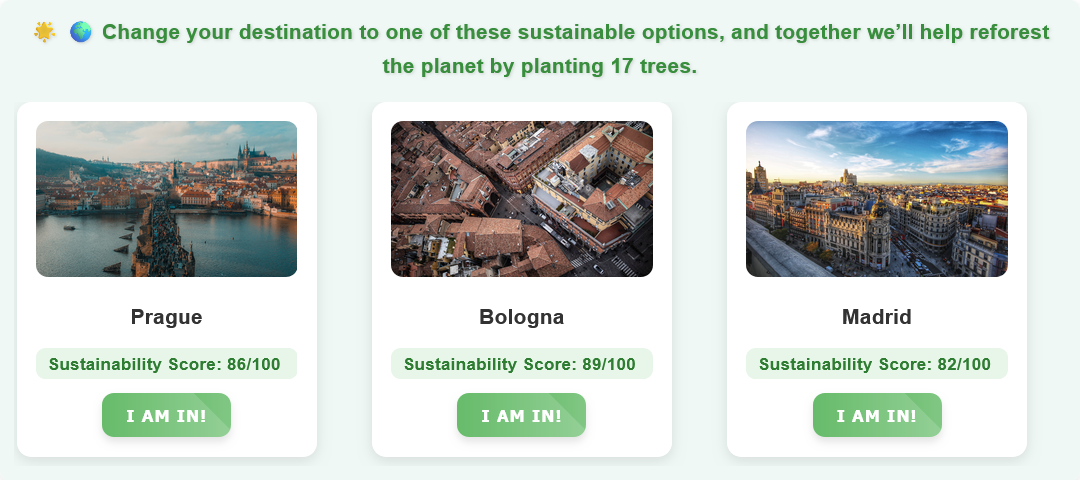}
    \caption{Alternative destination suggestions shown on the booking page when a high-emission option is selected. These alternatives are greener and align well with the user's interests, encouraging more sustainable choices.}
    \label{fig:alternatives}   
\end{figure}

\paragraph{Dynamic Trade-Off Nudges}  
When a chosen option is flagged as high-impact, the system deploys nudges to encourage sustainable behavior without dismissing user intent. Examples include:  
\begin{itemize}
    \item \textbf{Alternative Destination Banners} (\autoref{fig:alternatives}): Suggest greener but interest-aligned destinations.  
    \item \textbf{Positive Reinforcement}: For sustainable choices, highlight CO\textsubscript{2} savings (e.g., trees saved).  
\end{itemize}  
These banners appear contextually, on the Explore page (e.g., as highlighted in \textcolor{red}{red} in~\autoref{fig:card_view_single}), during booking as alternative suggestions (\autoref{fig:alternatives}), or at confirmation, framing trade-offs in terms of interest alignment and sustainability gains. The different banners are highlighted in~\autoref{fig:banners}.

\paragraph{Visual Encoding and Animations}  
Color coding facilitates rapid interpretation: green icons denote low-emission options, while red icons indicate high-impact choices, similar to the traffic light color coding scheme~\cite{magazine2020ux}. 
Labels such as “Best Match,” “Green Runner-up,” or “Eco-Bronze Medallist” provide comparative cues. 
We also use real-time animations, such as CO\textsubscript{2} progress bars or tree icons, to visualize emission thresholds and make trade-offs tangible, creating an engaging and informative user experience.

\paragraph{Journey Culmination}  
The experience concludes with a streamlined confirmation or booking flow (\autoref{fig:booking_page}), combining usability with subtle nudging mechanisms to ensure both user satisfaction and awareness of sustainability.

\section{System Design and Validation} \label{section: implementation}

In this section, we detail the core recommendation logic, provide an overview of the system implementation, and present a preliminary user study evaluating the effectiveness of usability and sustainability.

\subsection{Core Recommendation Logic} \label{subsection: scoring}

Destinations are ranked using a linear, interpretable score where lower values indicate better outcomes, preserving a consistent “less is better” interpretation across sustainability criteria. 
Our overall scoring function (Sustainability Score) is inspired from~\citet{banerjee2024modeling}, where the authors use a weighted sum of multiple factors to rank travel destinations and assign a sustainability indicator to the destination.
The score combines transport efficiency, popularity (in terms of the number of user-generated content, i.e., ratings and reviews at the destination), seasonality (defined in terms of the estimated crowdedness of the destination during the intended travel month), and interest alignment, along with personalization that reflects user-prioritized sustainability attributes such as air quality, climate vulnerability, and walkability. When personalization is absent, transport and interest alignment are automatically emphasized. 
For the start weights, we use the same weights as in the original work.

However, the weights are adaptable; for instance, if the user forgoes personalization, the weight of the interest match score is increased. A lower final score signifies a better overall recommendation. This composite metric enables a default ranking, while the UI components break it down to make the individual trade-offs visible.

\subsection{Implementation}
We implement this as a client-first, modular \textbf{React}\footnote{\url{https://react.dev}} application that centralizes state management via \textbf{Zustand}\footnote{\url{https://zustand.docs.pmnd.rs/getting-started/introduction}}, hosted on \textbf{Firebase}\footnote{\url{https://firebase.google.com/}}. 
User interactions and consented activity are logged asynchronously in \textbf{Firebase Firestore} database, 
while a session-local recommendation pipeline operates over a pre-computed publicly available dataset sourced from HuggingFace\footnote{\url{https://huggingface.co/datasets/ashmib/SynthTRIPs/blob/main/kb/eu-cities-database.csv}}. 
The frontend leverages \textbf{Material-UI}\footnote{\url{https://mui.com/material-ui}} for consistent styling and responsive design, while Leaflet.js powers the interactive map visualizations.

\subsection{Preliminary User Study}
\begin{figure}
    \centering
    \includegraphics[width=0.6\linewidth]{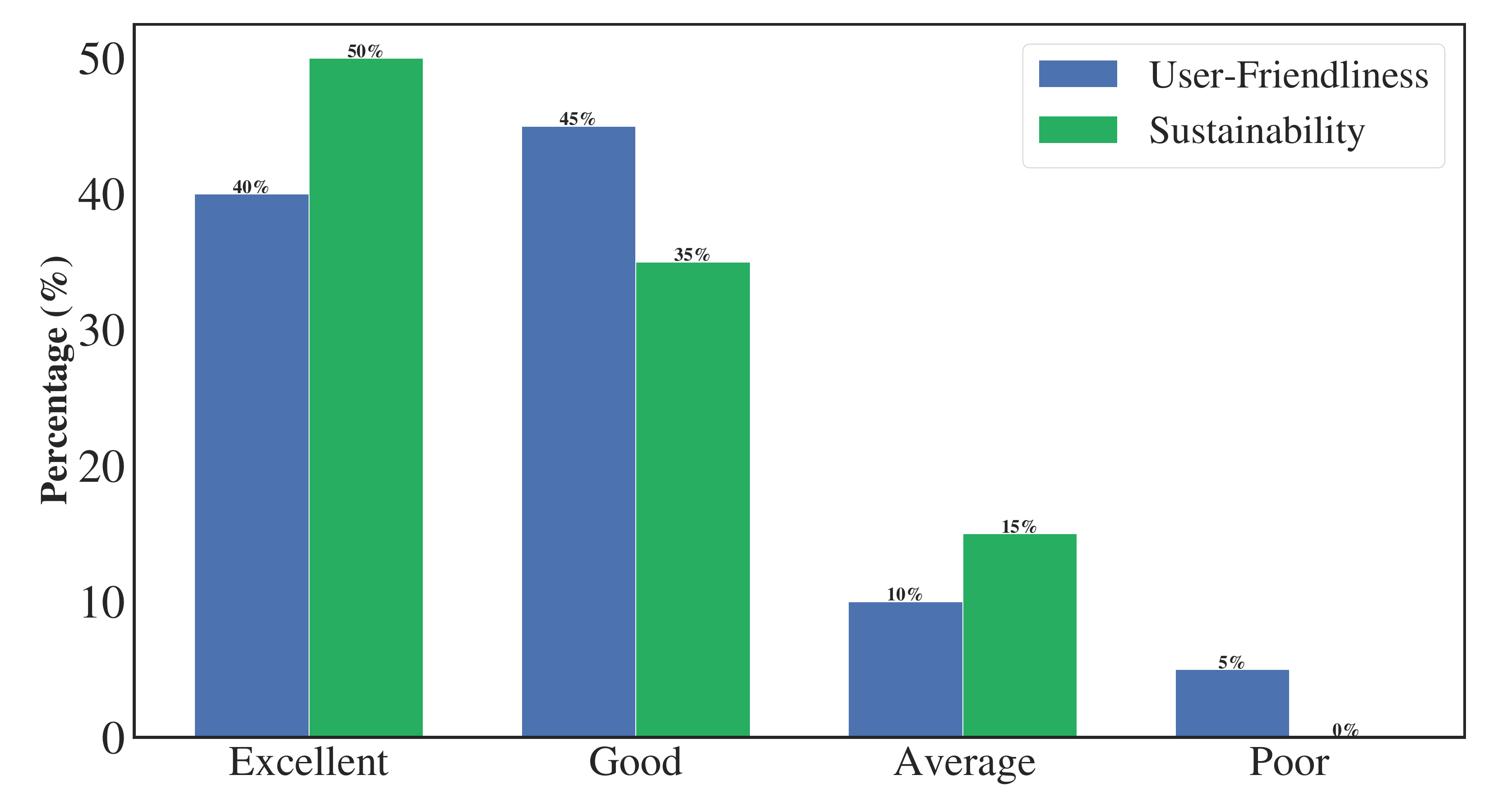}
    \caption{User feedback on the SmartSustain Recommender.}
    \label{fig:user_study}
\end{figure}

We conducted a preliminary user study with 21 participants to evaluate the usability and effectiveness of the SmartSustain Recommender. Participants explored the application after a brief introduction and provided feedback via a survey. They assessed both the overall user-friendliness and the system's effectiveness in promoting sustainable travel choices.

As shown in~\autoref{fig:user_study}, the system was perceived as highly usable, with 45\% of participants rating it “Good” and 40\% “Excellent.” Only 15\% rated it as “Average” or “Poor.” Regarding sustainability, 50\% of respondents considered the system highly effective at promoting eco-friendly choices, 35\% somewhat effective, and 15\% were neutral. No participants judged it ineffective. These results highlight the system's strengths in usability and sustainability, while indicating opportunities to refine features that enhance engagement.

Users particularly appreciated the clear visualization of trade-offs and contextual nudges. However, some features, such as advanced personalization options and the radar chart illustrating transport mode trade-offs, were less intuitive. Based on this feedback, we plan to iterate on the design to improve interface intuitiveness and provide clearer explanations of sustainability trade-offs, further enhancing the user experience.

\section{Conclusion} \label{section: future}
The SmartSustain Recommender serves as a powerful demonstration of how user-centric design can be applied to the complex problem of sustainable travel planning. 
By focusing on the clear visualization of trade-offs, implementing context-aware nudges, our system empowers users to make more informed and eco-conscious decisions. 
A preliminary user study indicated strong user satisfaction and perceived effectiveness with the system's features.

Future work will focus on integrating real-time data from transportation and accommodation APIs, expanding the dataset to include global destinations with finer-grained city attributes, and conducting large-scale user studies to quantitatively assess the impact of different visualization and nudging strategies on sustainable decision-making.

\section*{GenAI Usage Disclosure}
We used ChatGPT (OpenAI), Claude (Anthropic), and Gemini (Google) for code suggestions, and Grammarly for language refinement. All outputs were critically reviewed and revised, and we take full responsibility for the final content.

\bibliography{main}


\end{document}